\documentclass[prb,twocolumn,showpacs,superscriptaddress]{revtex4}
\usepackage[dvips]{epsfig}
\usepackage{graphicx}
\usepackage{amsmath,amssymb,lscape,float}
\usepackage{bm}
\usepackage{bbm}
\usepackage{dsfont}
\usepackage{longtable}

\newcommand{\norm}[1]{\lVert#1\rVert}
\renewcommand{\Re}{\mathop{\rm Re}\nolimits}
\newcommand{\hc}{\mathop{\rm H.c.}\nolimits}

\newcommand{\dg}{\dagger}

\newcommand{\gpsi}{{\psi}}

\newcommand{\gphi}{{\phi}}

\newcommand{\gvphi}{{\varphi}}

\newcommand{\mbf}{\mathbf}
\renewcommand{\vec}{\mbf}
\newcommand{\veck}{\vec k}
\newcommand{\vecr}{\vec r}

\makeatletter
\renewcommand\appendix{\par
  \setcounter{section}{0}%
  \setcounter{subsection}{0}%
  \setcounter{equation}{0}%
  \gdef\thesection{\Alph{section}}
  \@addtoreset {equation}{section}
  \renewcommand{\theequation}{\thesection\arabic{equation}}}
\makeatother

\begin{document}

\title{Electron-phonon coupling in graphene antidot lattices:\\
        an indication of polaronic behavior}

\author{Nenad Vukmirovi\'c}
\affiliation{Computational Research Division,
Lawrence Berkeley National Laboratory,
Berkeley, California 94720, USA}

\author{Vladimir M. Stojanovi\'c}
\email{vladimir.stojanovic@unibas.ch}
\affiliation{Department of Physics, University of Basel,
 Klingelbergstrasse 82, CH-4056 Basel, Switzerland}

\author{Mihajlo Vanevi\'c}
\affiliation{School of Physics, Georgia Institute of Technology,
Atlanta, Georgia 30332, USA}
\affiliation{Kavli Institute of Nanoscience, Delft University of Technology,
2628 CJ Delft, The Netherlands}

\date{\today}
\begin{abstract}
We study graphene antidot lattices -- superlattices of perforations (antidots) 
in a graphene sheet -- using a model that accounts for the phonon-modulation of the $\pi$-electron 
hopping integrals. We calculate the phonon spectra of selected antidot 
lattices using two different semi-empirical methods. Based on the adopted model, we quantify the nature 
of charge carriers in the system by computing the quasiparticle weight due to the electron-phonon interaction 
for an excess electron in the conduction band. We find a very strong phonon-induced renormalization, with the effective electron masses 
exhibiting nonmonotonic dependence on the superlattice period for a given antidot diameter. 
Our study provides an indication of polaronic behavior and points to the necessity of taking 
into account the inelastic degrees of freedom in future studies of transport in graphene antidot lattices. 
\end{abstract}
\pacs{63.20.kd, 63.22.-m, 71.38.-k, 73.21.Cd}
\maketitle

Recent years have seen a surge of interest in graphene -- the two-dimensional form
of carbon with atoms ordered in a honeycomb lattice.~\cite{GrapheneReviews}
This material shows extraordinary properties, such as 
room-temperature ballistic transport on a submicron scale and the 
possibility of heavy doping without altering significantly the charge-carrier 
mobility. Yet, the usefulness of pure graphene for carbon-based 
electronics~\cite{CarbonElectr} is limited as the electron transmission 
probability across a potential barrier is always unity -- regardless of 
the height and width of the barrier -- a feature akin to Klein 
tunneling.~\cite{KleinTunneling} Thus the conductivity cannot be altered 
by a gate voltage, the latter being a key property of a field-effect 
transistor.  

Motivated in part by the compelling need to create a band gap in 
graphene, an extensive research effort is currently being dedicated
to understanding the electronic properties of graphene-based 
superlattices.~\cite{GrapheneSuperlatt} A class of such structures,  
made by perforating a graphene sheet -- graphene {\em antidot
lattices} -- has recently been proposed.~\cite{Danci++:08} These 
lattices belong to the family of superhoneycomb systems~\cite{Shima+Aoki:93} 
and can be obtained by patterning graphene monolayers using electron-beam 
lithography, a method which allows feature sizes as small as tens of 
nanometers. It is worthwhile to stress, however, that -- owing to recent advances in 
nanofabrication~\cite{ExperProgressR:09} -- sub-10\:nm antidot
diameters constitute a realistic near-future prospect.

The electronic structure of triangular antidot lattices has been studied 
theoretically,~\cite{Vanevic++:09,Fuerst++:09} revealing 
features such as the existence of localized midgap states (flat- and quasi-flat bands). 
In addition, transport properties of their square-lattice counterparts have been investigated
experimentally,~\cite{AntidotExper} showing a transport gap and weak localization 
corrections to the conductance. 

In the present work, we study the influence of phonons on the electronic
properties of graphene antidot lattices. We calculate the phonon spectra of selected antidot
lattices using two independent semi-empirical methods. 
We then describe the electron-phonon (henceforth e-ph) interaction based on a model 
that accounts for the modulation of hopping integrals by the lattice displacements 
(Peierls-type e-ph coupling).~\cite{PeierlsCoupling} Within this model, 
we quantify the effect of phonons by computing the conduction-band quasiparticle 
spectral weight. We show that the phonon-induced renormalization is much stronger 
than in graphene itself, with the effective electron masses being typically in the range 
$3.7-5$ bare band masses.

\begin{figure}[b!]
\begin{center}
\epsfig{file=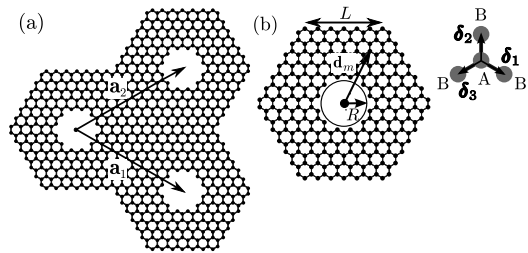,width=0.8\linewidth,clip=}
\caption{(a) A segment of a triangular graphene antidot lattice with circular antidots
and basis vectors $\mathbf{a}_{1}$ and $\mathbf{a}_{2}$. The lattice period is $|\mathbf{a}_{1}|=
|\mathbf{a}_{2}|=La\sqrt{3}$. (b) unit cell of an antidot lattice, with vectors 
$\bm{\delta}_{1},\:\bm{\delta}_{2}$, and $\bm{\delta}_{3}$ specifying 
positions of the nearest neighbors of a carbon atom on 
sublattice $A$.} \label{AntidotFig}
\end{center}
\end{figure}
The triangular graphene antidot lattice $\{L,R\}$ with circular perforations 
[see Fig.~\ref{AntidotFig}(a)] has a hexagonal unit 
cell [Fig.~\ref{AntidotFig}(b)] with side length $La$ and antidot radius 
$Ra$, where $a=2.46\:$\AA\: is the lattice constant of graphene. 
If we choose a carbon atom (hereafter C atom) on sublattice 
A as the origin, its nearest neighbors are given by the vectors 
$\bm{\delta}_{1}=(\sqrt{3}/2,-1/2)\:a_{cc}$, $\bm{\delta}_{2}=(0,1)\:a_{cc}$, 
and $\bm{\delta}_{3}=(-\sqrt{3}/2,-1/2)\:a_{cc}$ [Fig.~\ref{AntidotFig}(b)],
with $a_{cc}=1.42\:$\AA\:being the distance between adjacent C atoms.
 
Given the large size of unit cells in the antidot lattices that we consider -- with 
$N_{\textrm{at}}\sim 300-1600$ atoms -- the calculation of the electronic structure and
the phonon spectra in the whole Brillouin zone using {\em ab-initio} methods 
based on the density functional theory (DFT) is inconceivable. Instead, we model the 
band structure using a nearest-neighbor tight-binding Hamiltonian~\cite{Vanevic++:09} 
\begin{equation}
\hat{H}_{\textrm{e}}=-\frac{t}{2}\sum_{\mathbf{R},m,\bm{\delta}}
\big(\hat{a}^\dg_{\mathbf{R}+\mathbf{d}_{m}+\bm{\delta}}
\hat{a}_{\mathbf{R}+\mathbf{d}_{m}}+\hc\big) \:,
\end{equation}
\noindent where $\mathbf{R}$ designate the unit cells 
($N$ of them), $\mathbf{d}_{m}$ ($m=1,\ldots,N_{\textrm{at}}$) 
specify the positions of the C atoms within a unit cell, $\bm{\delta}$ stands for
the nearest neighbors of the C atom at position $\mathbf{R}+\mathbf{d}_{m}$,
and $t\approx 2.8$\:eV is the nearest-neighbor hopping integral. 
Within our model, the Bloch wave functions corresponding 
to the energy eigenvalues $\varepsilon_{n}(\veck)$ ($n$ is the band index) 
are given by $\gpsi_{n\mbf k}(\mbf r)=\sum_m 
C_m^{n,\veck}\gphi_{m\veck}(\vecr)$,
where $\gphi_{m\veck}(\vecr)=N^{-1/2}\sum_{\mbf R}
e^{i\veck \cdot \mbf R}\gvphi(\vecr -\mbf R -\mbf d_m)$ 
and $\gvphi(\vec r-\mbf R -\mbf d_m)$ is the $2p_z$-orbital of a
C atom at $\mbf R +\mbf d_m$. To a good approximation, the overlap of 
the $2p_z$-orbitals on different atoms can be neglected. 

\begin{figure}[t!]
\hspace*{-7.5mm}
\epsfig{file=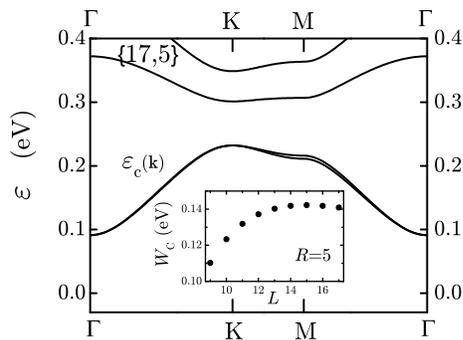,width=0.7\linewidth,clip=}
\caption{\label{CondDisper} The conduction-band dispersion $\varepsilon_{c}(\mathbf{k})$ 
for the $\{17,5\}$ antidot lattice. The inset shows the $L$-dependence ($9\leq L\leq 17$) 
of the conduction bandwidth $W_{c}$ for $R=5$.}
\end{figure}

The accuracy of the tight-binding method in the case of antidot lattices~\cite{Vanevic++:09} 
is corroborated by the recently demonstrated good agreement with the DFT results
for lattices with very small unit cells.~\cite{Fuerst++:09} Given that the underlying  
honeycomb lattice is bipartite, the resulting tight-binding energy spectrum has particle-hole 
symmetry~\cite{Vanevic++:09} -- a property not retained in the exact 
band structure.~\cite{Fuerst++:09} Our calculations show (see Fig.~\ref{CondDisper} for 
an illustration) that the antidot lattices are extremely narrow-band 
systems: for instance, in the $\{L,5\}$ family (with $9\leq L\leq 17$)
of lattices the conduction bandwidth $W_{c}$ increases from $0.11$\:eV
to $0.14$\:eV (see the inset of Fig.~\ref{CondDisper}); in the $\{L,7\}$ 
family ($12\leq L\leq 17$) it increases from $0.020$\:eV to $0.035$\:eV. 
The band gap decreases from $0.74$\:eV to $0.18$\:eV 
in the $\{L,5\}$ and from $0.30$\:eV to $0.15$\:eV in the $\{L,7\}$ family. 

The phonon spectrum of graphene was studied extensively, using either 
{\em ab-initio} methods or effective models.~\cite{GraphenePhononCalc}
In the present work, we calculate the phonon spectra of graphene {\em antidot lattices} using two independent 
methods that have recently been shown to yield very accurate results for graphene: the fourth-nearest-neighbor
force-constant (4NNFC) method, in the parametrization of Zimmermann {\em et al.}, 
and the valence force field (VFF) method of Perebeinos and Tersoff.~\cite{GraphenePhononCalc}

We study the phonon spectra for the $\{L,5\}$ and $\{L,7\}$ antidot lattices. 
In each case, we first find the equilibrium lattice configuration
by relaxing the atoms until forces on them are smaller than $10^{-5}$ eV/\AA. 
We then construct the force-constant matrix $D_{m\beta, m'\beta'} 
(\mbf{R}-\mbf{R}')\equiv\partial^2 E_{\textrm{tot}}/
\partial u_{m\beta}(\mbf{R})\partial u_{m'\beta'}(\mbf{R}')$,
where $u_{m\beta}(\mathbf{R})$ are the displacements ($\beta=x,y,z$) from the 
equilibrium position for an atom at $\mathbf{R}+\mbf d_m$, and $E_{\textrm{tot}}$
the total lattice potential energy. The normal-mode frequencies $\omega_{\lambda}(\mathbf{q})$ 
and eigenvectors $\mathbf{v}^{\lambda}(\mathbf{q})$ 
($\lambda$ is the phonon branch index) are obtained from
the eigenvalue problem $\mbf{D}(\mathbf{q})\mathbf{v}^{\lambda}(\mathbf{q})
=M\omega^{2}_{\lambda}(\mathbf{q})\mathbf{v}^{\lambda}(\mathbf{q})$ 
for the dynamical matrix $\mbf{D}(\mathbf{q})\equiv
\sum_{\mbf{R}}\mbf{D}(\mbf{R})\:e^{-i\mathbf{q}\cdot\mathbf{R}}$, 
with $M$ being the C-atom mass. 

\begin{figure}[b!]
\hspace*{-7.5mm}
\epsfig{file=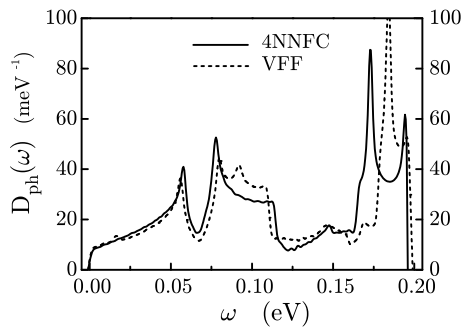,width=0.7\linewidth,clip=}
\caption{\label{PhononDOS} The phonon density-of-states for
the $\{17,5\}$ antidot lattice, obtained using the 4NNFC (solid line) 
and VFF (dashed line) methods.}
\end{figure}

The salient feature of the obtained phonon spectra is that the highest
optical-phonon energy at $\mathbf{q}=0$ is at around $195.3$\:meV --
essentially inherited from graphene itself and only weakly dependent on $L$ 
and $R$ -- while the lowest optical-phonon energy can be as low as
$0.69$\:meV, the case of the $\{17,5\}$ antidot lattice. 
The two methods used are compared by computing 
the phonon density-of-states $D_{\textrm{ph}}(\omega)\equiv N^{-1}\sum_{\mathbf{q},\lambda}
\delta[\omega-\omega_{\lambda}(\mathbf{q})]$, which shows their good 
agreement (Fig.~\ref{PhononDOS}), especially in the low-energy part of the 
phonon spectrum.

Generally speaking, the dominant mechanism of the e-ph coupling in the 
$\pi$-electron systems is the phonon-modulation of the electronic hopping 
integrals~\cite{SSH++,MahanEtAl,NonlocalCoupling}
-- Peierls-type coupling.~\cite{PeierlsCoupling} The latter forms the basis 
of the Su-Schrieffer-Heeger (SSH) model.~\cite{SSH++,SSHnanotubes}
We thus adopt a model comprising an electron term 
($\hat{H}_{\textrm{e}}$), the phonon term ($\hbar=1$) 
$\hat{H}_{\textrm{ph}}=\sum_{\mathbf{q},\lambda}
\omega_{\lambda}(\mathbf{q})(\hat{b}^{\dagger}_{\mathbf{q},\lambda}
\hat{b}_{\mathbf{q},\lambda}+1/2)$, and a Peierls-type e-ph coupling term
\begin{multline}\label{ephcoupling}
\hat{H}_{\textrm{ep}}=\frac{\alpha}{2}\sum_{\mathbf{R},m,\bm{\delta},\lambda}
\big(\hat{a}^\dg_{\mathbf{R}+\mathbf{d}_{m}+\bm{\delta}}
\hat{a}_{\mathbf{R}+\mathbf{d}_{m}}+\hc\big) \\
\times\big[\hat{\mathbf{u}}_{\lambda,\mathbf{R}+\mathbf{d}_{m}+\bm{\delta}}
-\hat{\mathbf{u}}_{\lambda, \mathbf{R}+\mathbf{d}_{m}}\big]\cdot 
\bar{\bm{\delta}} \:,
\end{multline}
\noindent where $\bar{\bm{\delta}}\equiv\bm{\delta}/\norm{\bm{\delta}}$ 
is the unit vector in the direction of $\bm{\delta}$,
\begin{equation}
\hat{\mathbf{u}}_{\lambda,\mathbf{R}+\mathbf{d}_{m}}\equiv
\frac{1}{\sqrt{N}}\sum_{\mathbf{q}}\frac{e^{i\mathbf{q}\cdot\mathbf{R}}
(\hat{b}^\dg_{-\mathbf{q},\lambda}+\hat{b}_{\mathbf{q},\lambda})}
{\sqrt{2M\omega_{\lambda}(\mathbf{q})}}\:\mathbf{v}^{\lambda}_{m}(\mathbf{q})
\end{equation}
\noindent is the displacement of an atom at position $\mathbf{R}+\mathbf{d}_{m}$ due 
to the (optical) phonon branch $\lambda$, and $\alpha=5.27\:\textrm{eV/\AA}$ the e-ph coupling constant 
describing the linear dependence of the $\pi$-electron hopping integrals upon the C-C bond-length 
modulations.~\cite{MahanEtAl} In the last equation, the phonon eigenvectors 
$\mathbf{v}^{\lambda}_{m}(\mathbf{q})$ are normalized such that 
$\sum_{m}\mathbf{v}^{\lambda}_{m}(\mathbf{q})\cdot[\mathbf{v}^{\lambda'}_{m}
(\mathbf{q})]^{*}=N_{\textrm{at}}\delta_{\lambda\lambda'}$ and 
$\mathbf{v}^{\lambda}_{m}(-\mathbf{q})=[\mathbf{v}^{\lambda}_{m}
(\mathbf{q})]^{*}$.~\cite{landau_statmech}

In momentum space, the Hamiltonian of Eq.~\eqref{ephcoupling} reads
\begin{equation}\label{mscoupling}
\hat{H}_{\textrm{ep}}=\frac{1}{\sqrt{N}}\sum_{\mathbf{k,q},\lambda,n}
\gamma_{nn}^{\lambda}(\mathbf{k,q}) \:
\hat{a}_{n,\mathbf{k+q}}^{\dagger}\hat{a}_{n,\mathbf{k}}
(\hat{b}_{-\mathbf{q},\lambda}^{\dagger}+\hat{b}_{\mathbf{q},\lambda}) \:,
\end{equation}
\noindent where $\hat{a}_{n,\mathbf{k}}^{\dagger}$ creates an electron in 
a Bloch state $\gpsi_{n\mbf k}$ (eigenstate of $\hat{H}_{\textrm{e}}$) and
$\gamma_{nn}^{\lambda}(\mathbf{k,q})$ stands for the e-ph interaction 
vertex function. It can be shown that the latter is given by 
$\gamma_{nn}^{\lambda}(\mathbf{k,q})=
V_{nn}^\lambda (\mbf k ,\mbf q)+W_{nn}^\lambda(\mbf k ,\mbf q)$, where
\begin{multline}\label{vertex1}
V_{nn}^\lambda (\mbf k ,\mbf q)=\frac{\alpha}{\sqrt{8M\omega_\lambda(\mbf q)}}
\sum_{m,\bm{\delta}}\bar{\bm{\delta}}\cdot [\mbf v^\lambda_{m+\delta}
(\mbf q)-\mbf v^\lambda_m(\mbf q)]\\
\times\big[(C^{n, \mbf k+\mbf q}_{m+\delta})^* C^{n,\mbf k}_m+ 
(C^{n, \mbf k + \mbf q}_m)^* C^{n, \mbf k}_{m+\delta}\big] 
\end{multline}
\noindent is the contribution due to hopping within 
a single unit cell (the indices $m+\delta$ denote neighbors 
$\mathbf{d}_{m}+\bm{\delta}$ of site $\mathbf{d}_{m}$ and the 
coefficients $C^{n,\mbf k}_{m}$ originate from the aforementioned 
tight-binding band-structure calculation), while
\begin{multline}\label{vertex2}
W_{nn}^\lambda (\mbf k ,\mbf q)=\frac{\alpha}{\sqrt{8M\omega_\lambda(\mbf q)}}
\sideset{}{'}\sum_{m,\bm{\delta},\mbf{a}}\bar{\bm{\delta}}
\cdot[e^{i\mbf q\cdot\mbf a}\mbf v^\lambda_{m_1}(\mbf q)-\mbf v^\lambda_m(\mbf q)]\\
\times\big[ e^{-i(\mbf k+\mbf q)\cdot \mbf a}  
(C^{n, \mbf k+\mbf q}_{m_1})^* C^{n,\mbf k}_m+ e^{i\mbf k\cdot \mbf a}  
(C^{n, \mbf k + \mbf q}_m)^* C^{n, \mbf k}_{m_1}\big]
\end{multline}
\noindent originates from the hopping between adjacent unit
cells. The prime in the last sum signifies a summation restricted to
the neighbors $\mbf d_m+\bm{\delta}$ of site $\mbf d_m$ that satisfy 
the condition $\mbf d_m + \bm \delta =\mbf a + \mbf d_{m_1}$
for some $m_1=m_1(\bm{\delta})$, with $\mbf a = \pm \mbf a_1$, $\pm \mbf a_2$, 
$\pm(\mbf a_1-\mbf a_2)$. Unlike the more conventional Holstein-type e-ph
coupling, which is completely momentum-independent, 
the Peierls-type coupling depends on both the electron and phonon 
momenta. The momentum-dependence of the vertex function
is more complicated than that of the standard SSH-coupling: while the 
latter is defined on a monoatomic lattice, here we study a 
lattice with a basis $\{\mathbf{d}_{m}\}$. It is straightforward to check, 
however, that for the case of a monoatomic lattice 
($N_{\textrm{at}}\rightarrow 1,\:C^{n,\mbf k}_{m}\rightarrow\delta_{nm}$) 
the vertex function takes on the standard SSH dependence $\gamma(\mbf{k},\mbf{q})
\propto[\sin(\mbf{k}\cdot\mbf{a})-\sin\big((\mbf{k+q})\cdot\mbf{a}\big)]$.~\cite{SSH++}

The overlap of the bare-electron Bloch state $\hat{a}^{\dagger}_{n\mathbf{k}}|0\rangle$
and the Bloch state $|\Psi_{n\mathbf{k}}\rangle$ of the coupled e-ph system  
defines the quasiparticle spectral weight 
$Z_{n}(\mathbf{k})\equiv|\langle\Psi_{n\mathbf{k}}|\hat{a}^{\dagger}_{n\mathbf{k}}|0\rangle|^{2}$, 
a quantity characterizing the renormalization of the electron Green's function by the 
e-ph interaction. Its inverse is given by~\cite{MahanBook} 
$Z^{-1}_{n}(\mathbf{k})=1-\partial_{\omega}\Re\Sigma_{n}
(\mathbf{k},\omega)|_{\omega=\varepsilon_{n}(\mathbf{k})}$,
where $\Sigma_{n}(\mathbf{k},\omega)$ is the self-energy
due to the e-ph coupling for an electron in the $n$-th Bloch band. 
Using the ordinary Rayleigh-Schr\"{o}dinger perturbation 
theory, one obtains 
\begin{equation}\label{sigmann}
\Sigma_{n}(\mathbf{k},\omega)=\frac{1}{N}\sum_{\mathbf{q},\lambda}
\frac{|\gamma_{nn}^{\lambda}(\mathbf{k,q})|^{2}}{\omega-\omega_{\lambda}
-\varepsilon_{n}(\mathbf{k+q})+i0^{+}} \:,
\end{equation}
\noindent where the dispersion of the optical phonons is neglected
in the denominator of the last expression. 
In particular, the Rayleigh-Schr\"{o}dinger perturbation theory is known to describe 
the properties of coupled e-ph systems at $\mathbf{k}=0$ better than the 
self-consistent Born approximation.~\cite{MahanBook}

In what follows, we compute the quasiparticle spectral weight due to the 
e-ph interaction for an electron at the bottom ($\mathbf{k}=0$) of the 
conduction ($n\rightarrow c$) band. From Eq.~\eqref{sigmann}, 
for this special case we obtain 
\begin{equation}\label{Zcinverse}
Z_{c}^{-1}(0)=1+\frac{1}{N}\sum_{\mathbf{q},\lambda}
\frac{|\gamma_{cc}^{\lambda}(\mathbf{k=0,q})|^{2}}{\big[\varepsilon_{c}
(0)-\varepsilon_{c}(\mathbf{q})-\omega_{\lambda}\big]^{2}} \:,
\end{equation}
\noindent a quantity that yields the ratio of the effective (in the presence of
the e-ph interaction) and the bare band electron masses: 
$Z_{c}^{-1}(0)=m_{\textrm{eff}}/m^{*}_{\textrm e}$. 
Based on Eq.~\eqref{Zcinverse}, with the aid of Eqs.~\eqref{vertex1} and \eqref{vertex2}, 
we evaluate $Z_{c}^{-1}(0)$ for the $\{L,5\}$ ($9\leq L\leq 17$) 
and $\{L,7\}$ ($12\leq L\leq 17$) families of lattices. 
These demanding numerical calculations are performed via 
parallelization on multiple processors.

The calculations show a rather strong phonon-induced renormalization (see Table~\ref{TableOfZ}) 
compared to graphene,~\cite{GraphenePhononEffect} where $Z=0.93$ (at the Dirac points) 
or larger.~\cite{ParkZgraphene} Importantly, we find a very 
good agreement between the results obtained using the 4NNFC and VFF phonon spectra, with 
the effective electron masses $m_{\textrm{eff}}$ from 4NNFC being slightly larger in all 
the cases considered. As can be inferred from Table~\ref{TableOfZ}, for fixed $L$ 
the renormalization is larger for the structures with smaller antidot diameters, 
which squares with intuition. Another interesting feature that we find 
is a nonmonotonic $L$-dependence of $m_{\textrm{eff}}$ 
for given $R$, with minima for $L=13$ in the $\{L,5\}$ 
family and $L=15$ in the $\{L,7\}$ family of antidot lattices. 

\begin{table}
\caption{\label{TableOfZ} Calculated inverse quasiparticle weights (electron mass 
renormalization) $Z^{-1}_{c}(0)=m_{\textrm{eff}}/m^{*}_{\textrm e}$ for various antidot 
lattices $\{L,R\}$, based on the phonon spectra obtained using the 4NNFC and VFF methods, 
respectively.}
\begin{ruledtabular}
\begin{tabular}{r c c}
  & \underline{\:$Z^{-1}_{c}(0)$ \:4NNFC\:} & \underline{$Z^{-1}_{c}(0)$\:\:\:\:VFF\:} \\
 L & R=5\qquad R=7 &  R=5\qquad R=7\\ \hline
 9 & \hspace{-0.7cm}$ 5.046$\quad\:\:\: $ $ &  \hspace{-0.7cm}$4.811$\quad\:\:\:  $ $ \\
10 & \hspace{-0.7cm}$4.836$\quad\:\:\:  $ $ &  \hspace{-0.7cm}$4.613$\quad\:\:\:  $ $ \\
11 & \hspace{-0.7cm}$4.732$\quad\:\:\:  $ $ &  \hspace{-0.7cm}$4.509$\quad\:\:\:  $ $ \\
12 & $4.681$\quad\:\:\:  $4.056$ &  $4.452$\quad\:\:\:  $4.030$ \\
13 & $4.662$\quad\:\:\:  $3.827$ &  $4.450$\quad\:\:\:  $3.808$ \\
14 & $4.668$\quad\:\:\:  $3.739$ &  $4.452$\quad\:\:\:  $3.725$ \\
15 & $4.684$\quad\:\:\:  $3.707$ &  $4.459$\quad\:\:\:  $3.699$ \\
16 & $4.709$\quad\:\:\:  $3.758$ &  $4.475$\quad\:\:\:  $3.701$ \\
17 & $4.733$\quad\:\:\:  $3.795$ &  $4.494$\quad\:\:\:  $3.756$
\end{tabular}
\end{ruledtabular}
\end{table}

Detailed analysis shows that the low-energy phonons (below $30$ meV) 
contribute at most $20$ percent of the overall spectral weight, 
while among the high-energy ones the largest contributions come from 
two narrow intervals, around $173$ meV and $194$ meV, respectively. 
These high-energy modes typically provide $75-80$ percent of the 
spectral weight and their salient feature is that they do not involve significant
atomic displacements in the vicinity of the antidot edges.

The obtained strong mass renormalization $m_{\textrm{eff}}/m^{*}_{\textrm e}=3.7-5$ 
suggests that the charge carriers in the system acquire polaronic character. Indeed, it 
is plausible to have polaronic charge carriers in a narrow-band system with a strong e-ph 
coupling -- a common situation in organic semiconductors;~\cite{NonlocalCoupling} compared 
to the latter, graphene antidot lattices have yet narrower conduction bands and lower 
dimensionality. Given that the system at hand -- due to its size and complexity -- is 
out of reach of the exact-diagonalization methods, we have utilized a perturbative 
approach. Thus the obtained results are not expected to hold quantitatively, but they 
should still be qualitatively valid. These results underscore the relevance of phonons 
in antidot lattices and show that transport in these systems, unlike in graphene, cannot 
be treated as purely ballistic; i.e., it ought to be modelled by taking into account the 
inelastic degrees of freedom.

It is appropriate to comment on the robustness of our results
for realistic antidot lattices, which may contain hydrogen (H)-terminated edges.
The H-atoms couple only to the $\sigma$-orbitals, while all the bands close to 
the Fermi energy originate from the $\pi$-electron states. These atoms give rise to a small
change of hopping integrals near an edge due to the ensuing 
geometrical relaxation, leading to a minor band-gap modification.~\cite{Fuerst++:09} 
Likewise, the influence of a handful of H-atoms on the dynamics of realistic antidot lattices 
is also not expected to be drastic, since -- as shown in the present work -- the dominant 
phonon modes involve only very small atomic displacements 
in the vicinity of edges. Therefore, while related issues certainly 
merit further investigation, our results are expected to be largely unaffected 
in realistic lattices.

In summary, we have studied the influence of phonons on the electronic properties 
of graphene antidot lattices. We have computed the phonon spectra for representative 
antidot lattices and determined the quasiparticle spectral weight due to the electron-phonon 
interaction for an electron at the bottom of the conduction band. We have shown that 
the phonon-induced renormalization in these narrow-band systems is much stronger than 
in graphene itself, providing an indication of the polaronic nature of charge carriers. 
Our study paves the way for future investigations of the effect of phonons on the 
electronic and transport properties of graphene-based superlattices.

\begin{acknowledgments}
We thank C. Bruder for useful discussions. V.M.S. acknowledges 
financial support from the Swiss NSF and the NCCR Nanoscience.
\end{acknowledgments}

\bibliography{Graphene,polaronref}
\bibliographystyle{apsrev}

\end{document}